\documentclass[10pt,twocolumn,superscriptaddress,floats,showpacs,nobalancelastpage,longbibliography,prb]{revtex4-2}

\usepackage{array,longtable}
\newcolumntype{L}{>{\tiny $}p{0.33\columnwidth}<{$}}
\newcolumntype{M}{>{\scriptsize $}p{0.33\columnwidth}<{$}}
\newcolumntype{N}{>{\scriptsize $}p{0.43\columnwidth}<{$}}
\usepackage{amsmath}
\usepackage{amssymb}
\usepackage{amsfonts}
\usepackage{graphicx}
\usepackage{hyperref}
\usepackage{tabularx}

\usepackage{float}
\usepackage{graphics}
\usepackage[caption=false]{subfig}
\usepackage{tikz}
\usepackage{times}
\usepackage{color}

\newcommand*\circled[1]{\tikz[baseline=(char.base)]{
            \node[shape=circle,draw,inner sep=1.pt] (char) {#1};}}

\usepackage{ulem}
 
\allowdisplaybreaks[1]

\newif\ifhyper
\hypertrue
\ifhyper
\hypersetup{
   citecolor = {green},
   colorlinks = {true}, 
   urlcolor = {blue}, 
   linkcolor = {blue}
}
\fi

\begin{document}
\title{Programmable order by disorder effect and underlying phases through dipolar quantum simulators}

\author{Huan-Kuang Wu}
\affiliation{Department of Physics, Condensed Matter Theory Center and Joint Quantum Institute, University of Maryland, College Park, Maryland 20742, USA}

\author{Takafumi Suzuki}
\affiliation{Graduate School of Engineering, University of Hyogo, Hyogo, Himeji 670-2280, Japan}

\author{Naoki Kawashima}
\affiliation{Institute for Solid State Physics, University of Tokyo, Kashiwa, Chiba 277-8581, Japan}
\affiliation{Trans-scale Quantum Science Institute, The University of Tokyo, Bunkyo, Tokyo 113-0033, Japan}

\author{Wei-Lin Tu}
\email{weilintu@keio.jp}
\affiliation{Faculty of Science and Technology, Keio University, 3-14-1 Hiyoshi, Kohoku-ku, Yokohama 223-8522, Japan}

\date{\today}

\begin{abstract}
In this work, we study two different quantum simulators composed of molecules with dipole-dipole interaction through various theoretical and numerical tools.
Our first result provides knowledge upon the quantum order by disorder effect of the $S=1/2$ system, which is programmable in a quantum simulator composed of circular Rydberg atoms in the triangular optical lattice with a controllable diagonal anisotropy.
When the numbers of up spins and down spins are equal, a set of sub-extensive degenerate ground states is present in the classical limit, composed of continuous strings whose configuration enjoys a large degree of freedom.
Among all possible configurations, we focus on the stripe~(up and down spins aligning straightly) and kinked~(up and down spins forming zigzag spin chains) patterns.
Adopting the the real space perturbation theory, we estimate the leading order energy correction when the nearest-neighbor spin exchange coupling, $J$, is considered, and the overall model becomes an effective \textsl{XXZ} model with a spatial anisotropy.
Our calculation demonstrates a lifting of the degeneracy, favoring the stripe configuration.
When $J$ becomes larger, we adopt the infinite projected entangled-pair state~(iPEPS) and numerically check the effect of degeneracy lifting.
The iPEPS results show that even when the spin exchange coupling is strong the stripe pattern is still favored.
Next, we study the dipolar bosonic model with tilted polar angle which can be realized through a quantum simulator composed of cold atomic gas with dipole-dipole interaction in an optical lattice.
By placing the atoms in a triangular lattice and tilting the polar angle, the diagonal anisotropy can also be realized in the bosonic system.
With our cluster mean-field theory calculation, we provide various phase diagrams with different tilted angles, showing the abundant underlying phases including the supersolid.
Our proposal indicates realizable scenarios through quantum simulators in studying the quantum effect as well as extraordinary phases.
We believe that our results indicated here can also become a good benchmark for the two-dimensional quantum simulators.

\end{abstract}

\maketitle

\section{\label{Intro}Introduction}

The quantum many-body systems are often represented by Hamiltonian whose composing parts do not commute with each other.
This suggests that although the properties of low-energy states can be easily identified in a certain limit, often referred to as the classical limit, considering the whole Hamiltonian will make the estimation of ground state difficult because one needs to consider the superposition of every state in the full Hilbert space.
Moreover, the extra quantum or even thermal effect may lead to further stabilization among some competing low-energy states, causing the energy level crossing or degeneracy lifting, known as the order by disorder~(OBD) effect~\cite{J.Phys.41.1263, doi:10.1063/1.338570, ERastelli_1988, PhysRevLett.62.2056, PhysRevLett.64.88, AVChubukov_1991, PhysRevLett.68.855, QSheng_1992, PhysRevB.53.6455, PhysRevB.58.12049, PhysRevLett.85.3269, Nat.Phys.3.487, PhysRevLett.101.047201, PhysRevLett.109.077204, PhysRevB.89.014424, PhysRevLett.113.237202, PhysRevB.91.174424, PhysRevB.92.184416, PhysRevX.5.041035, PhysRevB.93.184408, PhysRevB.94.094438, PhysRevLett.121.237201}.
Besides, the introduction of frustration, through the Hamiltonian~\cite{Rep.Prog.Phys.80.016502, RevModPhys.89.025003, PhysRevB.96.045119, JPhysCondensMatter.32.455401} or lattice geometry~\cite{PhysRevLett.79.2554, Bramwell1495, Europhys.Lett.89.10010, NewJ.Phys.12.065025, Struck996, Rep.Prog.Phys.78.052502}, can also result in the many-fold degeneracy of ground state in the classical picture and such degeneracy enriches the quantum or thermal fluctuation.

Among all, the triangular lattice is the simplest lattice structure for inducing the geometrical frustration.
Particularly, the inclusion of anisotropy could lead to the formation of long strings composed of the same spins~\cite{JPhysCondensMatter.25.406003}.
The shape of strings enjoys a great degree of freedom, leading to a many-fold degeneracy.
Classically, such systems and their degeneracies can be realized, composed of self-assembled colloidal particles in a monolayer~\cite{Nature.456.898, PhysRevLett.102.048303}. 
For this soft-matter system, particles move to the opposite walls that confine them to maximize the free volume, corresponding to the up-and-down spin scenario and forming a good platform for studying the dynamics of frustration.
In $d$ dimensions, the entropy of degenerate ground states is proportional to $L^{d-1}$, with the number of sites $N\sim L^d$, resulting in a sub-extensive degeneracy. 
For the above-mentioned frustrated colloidal system, its thermal OBD effect has been studied and the authors unveiled the conclusion that the straight stripe should be favored~\cite{Shokef11804}.
On the other hand, the quantum counterpart for such OBD effect in zero temperature is also of interest.
When an extra non-commutable term, such as the spin exchange coupling, is introduced to the Ising-like Hamiltonian, the quantum fluctuation could further lower the energy and the amount is dependent on the classical configurations, resulting in the lifting of degeneracy.
Starting from the identical classical system, usually the quantum and thermal fluctuations result in similar outcomes and select the same ground state for weakly frustrated magnets. However, this is not always guaranteed and many counter examples have been previously provided~\cite{PhysRevLett.101.047201, PhysRevLett.113.237202, PhysRevB.94.094438, PhysRevB.102.220405}.
Therefore, it is worthwhile to examine the quantum version of this OBD effect. However, this effect has not yet been studied to the best of our knowledge. 
One possible reason is because such system can hardly be prepared in real materials. 
However, the recent success in simulating various quantum systems with neutral Rydberg atoms~\cite{book.Stebbings, book.Gallagher} has attracted huge attention, leading to the so-called quantum simulator~\cite{RevModPhys.82.2313, Wu_2021, Science2021, Nat.607.667}.
It has been shown that the quantum Ising model with both longitudinal and transverse fields can be simulated~\cite{PNAS2021} by utilizing the Rydberg blockade~\cite{PhysRevLett.85.2208, PhysRevLett.87.037901}.
Other spin models, such as the quantum XY model~\cite{PhysRevLett.120.063601, Science2019, Nat.Pho.16.724}, XXZ model~\cite{PhysRevX.11.011011, PRXQuantum.3.020303}, or even the XYZ model~\cite{Science2021_2}, can also be generated through the microwave engineering.
In the following content we will provide a possible setup to study the quantum OBD effect in our target system. 
We also conduct both analytical and numerical calculations and conclude that the stripe pattern is ultimately selected by the OBD effect.
Our conclusion based on the theoretical analysis could become a good benchmark for the real-world device of a two-dimensional quantum simulator using Rydberg atoms in the future.

Moreover, thanks to the advance of cold-atom technique~\cite{RevModPhys.80.885, Nat.Phys.8.267, Rep.Prog.Phys.76.086401, RevModPhys.91.035001}, many physical scenarios can now be simulated with emphasized quantum effect, leading to another type of quantum simulators.
Especially, the dipolar quantum gases made of erbium and dysprosium recently attract wide attention for the realization of supersolid out of its Bose-Einstein condensate~\cite{PhysRevLett.122.130405, PhysRevX.9.011051, PhysRevX.9.021012, PhysRevLett.123.050402, Nature.574.382}. 
Note that, by placing dipolar molecules in optical lattices, researchers have realized suitable quantum simulators for various Hamiltonian with electric~\cite{Schachenmayer_2010, PhysRevLett.107.115301, PhysRevA.84.033619, Nat.501.521, PhysRevLett.113.195302} or magnetic dipoles~\cite{doi:10.1126/science.aac9812}.
While Rydberg atoms are often used to create the electric dipole-dipole interaction to synthesize the spin exchange interaction, for the molecules carrying magnetic dipoles it can be viewed as a bosonic system probed by the Bose-Hubbard model~\cite{Chomaz_2023, Nature622.724, arXiv2405.07775}.
This makes the related studies for various extended Bose-Hubbard models necessary and many works have previously been
done~\cite{PhysRevLett.104.125301, doi:10.1143/JPSJ.80.113001, PhysRevA.85.021601, PhysRevLett.112.127203, PhysRevA.86.063635, NewJPhys.17.123014, PhysRevA.102.053306, PhysRevA.103.043333, CommunPhys5.65, PhysRevA.105.053301, PhysRevA.105.063302, JLowTempPhys209.34, PhysRevA.106.043301, PhysRevA.107.043318, PhysRevB.108.134201}. 
In our previous work, we have studied the effect by tilting the polar angle of dipole and realized a competing scenario between different lattice structures~\cite{PhysRevA.102.053306}. 
Within the original square lattice, we discovered that a three sub-lattice supersolid phase ought to show up by tilting the polar angle.
Related studies have also demonstrated the potential of such techniques in fabricating artificial physical scenarios of interest.
Besides the quantum OBD effect by Rydberg-atom simulator, therefore, in this work we also study the hard-core bosons with dipole-dipole interaction in the triangular lattice with different tilted dipolar angles, mimicking the distortion of lattice.

In the following content, we will first study the quantum OBD effect using the real space perturbation theory~(RSPT)~\cite{JPhysCondensMatter.1.2857, PhysRevB.48.7227, PhysRevLett.113.237202, JPhysConfSer.592.012110, PhysRevLett.118.147204, PhysRevB.102.245102} and infinite projected entangled-pair state~(iPEPS)~\cite{PhysRevLett.101.250602, NatRevPhys.1.538, RevModPhys.93.045003}.
We first map the two-level Rydberg system to a spin-1/2 scenario and show that when the number of up spins~($n_{\uparrow}$) is equal to that of down spins~($n_{\downarrow}$), a sub-extensive degeneracy is present in the classical limit.
Introducing the spin exchange coupling, energy correction can be considered with RSPT, and we focus on the leading-order correction between the stripe and kinked configurations.
Our calculation shows that the energy correction favors the stripe pattern, coinciding with previous results for thermal OBD~\cite{Shokef11804}.
As a further confirmation we apply the iPEPS calculation for a simplified toy model, showing that the stripe state is indeed more stable.
Next, for the dipolar bosons in triangular optical lattice with tilted polar angle we conduct numerical simulations with the cluster mean-field theory~(CMFT), which has been shown effective in capturing distinct phases in our previous work~\cite{PhysRevA.102.053306}.
We unveil fruitful phase diagrams with various underlying phases, including the supersolid phase. 
We then conclude our work and argue some potential scopes for further studies.

\section{\label{QOBD}Quantum OBD effect}

\begin{figure*}[tbp]
\centering
 	\includegraphics[width=2.0\columnwidth]{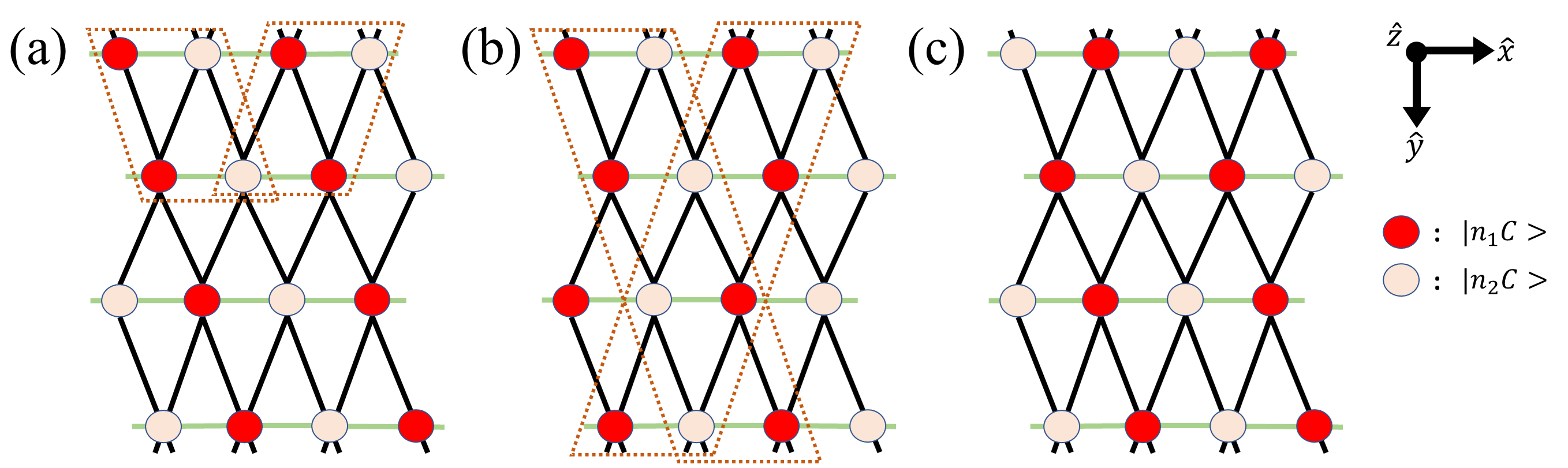}
\caption{Configurations for (a) stripe, (b) kinked, and (c) mixed states. Red~(pink) dots indicate that the Rydberg atom occupies the $|n_1C\rangle$~($|n_2C\rangle$) state, and the black~(green) bonds represent the bonds for diagonal~(horizontal) interaction, written as $J^z_d$~($J^z_{h}$) in Eq.~(\ref{HamiltonianT})~(also see the context). The brown dotted boxes indicate the minimum repeated unit cell for the stripe and kinked patterns, which we will later adopt for the iPEPS calculation.}
\label{fig1}
\end{figure*}

\subsection{\label{Rydberg}Rydberg-atom quantum simulator}

The adoption of Rydberg atoms for a quantum simulator has recently become a topic of interest.
It has been shown that the quantum Ising model can be synthesized in a large scale with many interesting properties~\cite{Nat551.579, PhysRevX.8.021070, Nat595.227}, as well as the potential application for quantum information~\cite{RevModPhys.82.2313}.
However, the adoption of Rydberg atoms for quantum simulation or computation
is largely hindered by the limited lifetime~(a few 100 \textmu s) for low-angular-momentum~($\ell$) Rydberg states~\cite{Saffman_2016}.
In contrast, when exciting the Rydberg states to a very large principal quantum number~($n$)
and utilizing the levels with largest $\ell$ and magnetic quantum number~($m$), the lifetime of these Rydberg states can be greatly enhanced, named after the circular Rydberg states~\cite{RevModPhys.73.565}.
Their naturally long lifetime originates from the fact that these large-$\ell$ states can only decay through the emission of a low-frequency photon, which is polarized parallel to the quantization axis. 
This feature makes the lifetime of circular Rydberg states much longer than that of regular low-$\ell$ Rydberg states.
Since it is a single-mode decay, the lifetime can be further elongated by placing the atoms in an environment which prohibits the field of resonance of the atomic transition~\cite{PhysRevLett.47.233, PhysRevLett.55.2137}.
Adopting a capacitor parallel to the plane where atoms reside, the lifetime again gets largely enhanced and an useful lifetime of at least 1.1 s for an atomic chain has previously been reported~\cite{PhysRevX.8.011032}.
Such measure can even be helpful for a non-zero-temperature environment and a millisecond-lived circular Rydberg state has recently been reported in room temperature~\cite{PhysRevLett.130.023202}.
This makes the quantum simulation and quantum computation much more feasible~\cite{PhysRevX.8.011032, PhysRevResearch.2.023192, PRXQuantum.2.030322}.

From now on we denote the circular Rydberg states as $|nC\rangle$ and such states fulfill the following condition: $\ell=|m|=n-1$.
These states have a wavefunction distribution of a torus whose circular orbit has a radius $r_n=a_0n^{3/2}$, where $a_0$ is the Bohr radius.
The circular states can be stabilized by an external electric field~($F$), which also defines the quantization axis, perpendicular to the plane of circular orbits.
Through a proper measure a two-level system composed of two circular states, $|n_1C\rangle$ and $|n_2C\rangle$ where $n_2>n_1$ and $C$ means the circular state, can be generated.

The interaction between any two Rydberg atoms comes from the electric dipoles carried by the atoms themselves, leading to a manifest dipole-dipole interaction~\cite{PhysRevA.77.032723, Shi_2022}. When two atoms are far from each other the interaction hardly affects the two-atom Rydberg state and thus a second-order perturbation is adopted, giving rise to a van der Waals interaction proportional to $1/r^6$, where $r$ represents the distance between sites~\cite{PhysRevA.75.032712, PhysRevA.96.032716}.
On the other hand, when the energy defect of two dipole-dipole coupled states is tuned to zero the strongest Rydberg-Rydberg interaction, proportional to $1/r^3$, appears through the Förster resonance process~\cite{Walker_2005, Nat.Phys.10.914}.
According to Ref.~\onlinecite{PhysRevX.8.011032} one needs to choose $n_2=n_1\pm2$ to have a flexible simulator, where both the spin-exchange and van der Waals interaction are proportional to $1/r^6$.
Since the spin-exchange effect might lead to the decay of circular states to nearby degenerate states, a magnetic field~($B$) parallel to the electric field is cast so that the highly degenerate hydrogenic manifold of the same principal quantum number $n$ can be lifted and the circular state is thus isolated~\cite{PhysRevX.8.011032}.
The magnetic field also prohibits the hybridization of spin up and down channel in each circular state. As a result, we obtain clean two-level Rydberg states for the purpose of quantum simulation.

The Hamiltonian for a pair of Rydberg atoms with the following basis, [$|n_1C,n_1C\rangle$, $|n_1C,n_2C\rangle$, $|n_2C,n_1C\rangle$, $|n_2C,n_2C\rangle$], can then be written as~\cite{PhysRevX.8.011032}
\begin{gather}
 H
 \propto
  \begin{bmatrix}
   h_{n_1-n_1} & 0 & 0 & 0\\
   0 & h_{n_1-n_2} & v_{n_1-n_2} & 0\\
   0 & v_{n_1-n_2} & h_{n_1-n_2} & 0\\
   0 & 0 & 0 & h_{n_2-n_2}
   \end{bmatrix},
\label{hmatrix}
\end{gather}
where in the Pauli basis it reads
\begin{equation}
\begin{aligned}
H\propto J(\sigma^x_1\sigma^x_2+\sigma^y_1\sigma^y_2)+J_z\sigma^z_1\sigma^z_2-B_z(\sigma^z_1+\sigma^z_2),
\end{aligned}
\label{HamiltonianR}
\end{equation}
with
\begin{equation}
\begin{aligned}
&J\propto|v_{n_1-n_2}|,\\
&J_z\propto h_{n_1-n_1}+h_{n_2-n_2}-2h_{n_1-n_2},\\
&B_z\propto \frac{1}{2}(h_{n_1-n_1}-h_{n_2-n_2}).
\end{aligned}
\label{Hparam}
\end{equation}
While $J$ is independent on the magnetic and electric fields, $J_z$ and $B_z$ are flexible varying the field strength and when the intersite spacing is $5\mu$m, $h_{48-48}=2.2$ GHZ$\mu$m$^6$, $h_{48-50}=2.66$ GHZ$\mu$m$^6$, and $h_{50-50}=3.03$ GHZ$\mu$m$^6$ when $F$=9 V/cm and $B$=13 Gauss, while $v_{48-50}=-0.539$ GHZ$\mu$m$^6$ stays independent of $F$ and $B$~\cite{PhysRevX.8.011032}.

Placing the circular Rydberg atoms in the optical lattice~\cite{PhysRevLett.124.123201, PhysRevA.101.013434}, a highly tunable simulator for the effective XXZ model with longitudinal field can thus be formed.
In order to engineer the triangular lattice to simulate the quantum OBD effect, a better way is to trap the individual atom through the optical tweezers~\cite{Natphys.17.1324}. 
One recent work has demonstrated such as possibility by trapping the rubidium circular Rydberg atoms into a $3\times6$ array of tweezers~\cite{PhysRevLett.131.093401}. 
It is reported that the trapping time can scale to over a few milliseconds while the simulation time only demands a few microseconds.
A very recent paper has reported the existence of long-lived circular Rydberg states by trapping the alkaline-earth atoms through the optical tweezers~\cite{PhysRevX.14.021024}. Their discovery also widens the potential usage of this kind of quantum simulator.
Therefore, by properly manipulating the positions of optical tweezers the deformed triangular optical lattice can be generated, as shown in Fig.~\ref{fig1}.
There, the diagonal bonds~(black bonds) is longer in its intersite spacing so that the sub-extensive degeneracy in the classical limit~($J=0$) can be simulated.
We demonstrate three different configurations in Fig.~\ref{fig1} for the (a) stripe, (b) kinked, and (c) mixed configurations.
Mixture of stripe and kinked configurations in a larger lattice gives rise to various possible states, forming the many-fold degeneracy.
Note that the diagonal spacing needs to be large enough to prohibit the off-diagonal magnetic orders.

While the basic setup has been introduced, next we will adopt the theoretical and numerical tools to study this quantum OBD effect, which can benchmark the real-world device of two-dimensional quantum simulator with Rydberg atoms.

\subsection{\label{RSPT}Real space perturbation theory}

We first introduce the RSPT and derive the formula in this section.
\if
Since Eq.~(\ref{HamiltonianR}) effectively represents an Ising model with both longitudinal and transverse fields, we can re-write the Hamiltonian with spin-1/2 operators
\begin{equation}
\begin{aligned}
&\frac{1}{2}\sum_{ij} |R\rangle\langle R|_i\otimes|R\rangle\langle R|_j\to \sum_{\langle ij\rangle} \hat{S}^z_i\hat{S}^z_j,\\
&\sum_{i}\frac{\Omega}{2}(|G\rangle\langle R|_i+|R\rangle\langle G|_i)-\delta|R\rangle\langle R|_i\to - \sum_i \textbf{B}\cdot\hat{\textbf{S}}_i.
\end{aligned}
\label{HamiltonianS}
\end{equation}
\fi
In this work we study the effect of spin-exchange quantum fluctuation to the system of Rydberg atoms, which can be represented as $\sum_{\langle ij \rangle}(\hat{S}^+_{i}\hat{S}^-_{j}+\hat{S}^-_i\hat{S}^+_j)$.
As a result, the target Hamiltonian represented by spin-1/2 operators can be written as
\begin{equation}
\begin{aligned}
H&=J\sum_{\langle ij \rangle}(\hat{S}^+_{i}\hat{S}^-_{j}+\hat{S}^-_i\hat{S}^+_j)+\sum_{\langle ij\rangle} J^z_{ij}\hat{S}^z_i\hat{S}^z_j- \sum_i \textbf{B}\cdot\hat{\textbf{S}}_i,
\end{aligned}
\label{HamiltonianT}
\end{equation}
where $J^z_{ij}\gg |J|, |\textbf{B}|$ and $\textbf{B}$ represents the external Zeeman field. We assign $J$ to be isotropic since later we will find out that only the perturbative processes concerning the diagonal spin exchange would matter.
To effectively reflect the anisotropy caused by the deformed lattice, we assign that $J^z_{ij}$ is equal to $J^z_d$ when the bond is in the diagonal direction~(black bonds in Fig.~\ref{fig1}) and $J^z_h$ when the bond is in the horizontal direction~(green bonds in Fig.~\ref{fig1}).
When $J^z_d<J^z_h$ and $n_{\uparrow}=n_{\downarrow}$, the ground state configuration enjoys a many-fold degeneracy because the same spins tend to align along the diagonal direction. 
However, due to the fact that there are two diagonal bonds, we can have many different degenerate configurations and the total number is equal to $2^{L_y-1}$, where $L_y$ is the number of rows along $\hat{y}$ direction. 
As a result, the ground state entropy $S_{GS}\sim L_y-1$, leading to a sub-extensive degeneracy. 
Among these degenerate configurations there are two special patterns, the stripe and kinked states. 
The stripe state is composed of parallel straight lines of up and down spins~(Fig.~\ref{fig1}(a)), while for the kinked state there are three sub-lattices occupied by up or down spins within each vertical diamond of the triangular lattice~(Fig.~\ref{fig1}(b)). 
The rest degenerate states, called the mixed states, are simply the combinations of stripe and kinked states in a specific proportion~(one example shown in Fig.~\ref{fig1}(c)).

For applying RSPT, one needs to rotate the local coordinate on each site so that every spin points along the $\hat{z}$-direction. 
Thus, we re-write the corresponding spin-1/2 Hamiltonian in the following form
\begin{equation}
\begin{aligned}
H=\sum_{\langle ij \rangle}\textbf{S}_i\cdot \textbf{J}\cdot \textbf{S}_j- \sum_i \textbf{B}\cdot \textbf{S}_i ,
\end{aligned}
\label{HamiltonianRSPT}
\end{equation}
with
\begin{equation}
\textbf{J}=\left\{
\begin{aligned}
&\text{diag}(2J,2J,J^z_{h})\;\;\;\text{if}\:\langle ij\rangle \parallel \text{horizontal direction}\\
&\text{diag}(2J,2J,J^z_{d})\;\;\;\text{if}\:\langle ij\rangle \parallel \text{diagonal direction}.
\end{aligned}
\right.
\label{J}
\end{equation}
%
We can extend this to a more generalized Hamiltonian with longer-range hopping or even off-diagonal terms if needed, simply by modifying the form of coupling matrices. 
In Eq.~(\ref{HamiltonianRSPT}), $\textbf{S}_i=S^+_i\hat{e}^-_i+S^-_i\hat{e}^+_i+S^z_i\hat{e}^z_i$ with $\hat{e}^\pm_i=\frac{1}{2}(\hat{e}^x_i\pm i\hat{e}^y_i)$ and the classical configuration on each site is $\textbf{S}^{(0)}_i=S\hat{e}^z_i$.
Note that now the $S^\pm$ and $S^z$ are no longer operators. They simply represent the corresponding operation for the classical spin.
Our choice of the ``unit" vectors guarantees the transformation to be unitary, although its length is not necessarily the unity. This also explains why we have a pre-factor of 2 for $J$ in Eq.~(\ref{J}). 
Next, we expand the equation and the result becomes
\begin{equation}
\begin{aligned}
H=&\sum_{\langle ij \rangle}(\hat{e}^+_i\cdot \textbf{J}\cdot\hat{e}^+_jS^-_iS^-_j+H.C.)\\
&\;\;+(\hat{e}^+_i\cdot \textbf{J}\cdot\hat{e}^-_jS^-_iS^+_j+H.C.)\\
&\;\;+(\hat{e}^+_i\cdot \textbf{J}\cdot\hat{e}^z_jS^-_iS^z_j+\hat{e}^z_i\cdot \textbf{J}\cdot\hat{e}^+_jS^z_iS^-_j+H.C.)\\
&\;\;+\hat{e}^z_i\cdot \textbf{J}\cdot\hat{e}^z_jS^z_iS^z_j\\
&-\sum_i \textbf{B}\cdot\hat{e}^-_iS^+_i+\textbf{B}\cdot\hat{e}^+_iS^-_i+\textbf{B}\cdot\hat{e}^z_iS^z_i .
\end{aligned}
\label{HamiltonianRSPT2}
\end{equation}
After introducing the quantum fluctuation, the spin configuration does no longer align along the local $\hat{e}^z_i$ direction and thus we can replace $S^z_i$ by $S-\delta S_i$. Then, Eq.~(\ref{HamiltonianRSPT2}) becomes
\begin{equation}
\begin{aligned}
H=H_0+H_{\text{unperturbed}}+H_{\text{p}},
\end{aligned}
\label{HamiltonianRSPT3}
\end{equation}
where
\begin{equation}
\begin{aligned}
H_0=S^2\sum_{\langle ij\rangle}\hat{e}^z_i\cdot \textbf{J}\cdot\hat{e}^z_j-S\sum_i\textbf{B}\cdot \hat{e}^z_i,
\end{aligned}
\label{Hamiltoniancla}
\end{equation}
representing the classical energy. And $H_{\text{unperturbed}}$ is
\begin{equation}
\begin{aligned}
H_{\text{unperturbed}}=&-S\sum_{\langle ij\rangle}\hat{e}^z_i\cdot \textbf{J}\cdot\hat{e}^z_j(\delta S_i+\delta S_j)\\
&+\sum_{\langle ij\rangle}\hat{e}^z_i\cdot \textbf{J}\cdot\hat{e}^z_j\delta S_i\delta S_j+\sum_i \textbf{B}\cdot \hat{e}^z_i\delta S_i.
\end{aligned}
\label{Hamiltonianunp}
\end{equation}
Since Eq.~(\ref{Hamiltonianunp}) does not change the spin configuration, it will not contribute to the perturbation. At last, the perturbative Hamiltonian $H_p$ is written as
\begin{equation}
\begin{aligned}
H_{\text{p}}=&\sum_{\langle ij\rangle}(J^{(1)}_{ij}+J^{(2)}_{ij}+J^{(3)}_{ij})-\sum_{i}B_i,
\end{aligned}
\label{Hamiltonianper}
\end{equation}
where
\begin{equation}
\begin{aligned}
&J^{(1)}_{ij}=\hat{e}^+_i\cdot \textbf{J}\cdot\hat{e}^+_jS^-_iS^-_j+H.C.\\
&J^{(2)}_{ij}=\hat{e}^+_i\cdot \textbf{J}\cdot\hat{e}^-_jS^-_iS^+_j+H.C.\\
&J^{(3)}_{ij}=\hat{e}^+_i\cdot \textbf{J}\cdot\hat{e}^z_jS^-_i S^z_j+\hat{e}^z_i\cdot \textbf{J}\cdot\hat{e}^+_j S^z_iS^-_j+H.C.\\
&B_i=\textbf{B}\cdot\hat{e}^-_iS^+_i+\textbf{B}\cdot\hat{e}^+_iS^-_i.
\end{aligned}
\label{V}
\end{equation}
With these transverse perturbative terms, we can evaluate the energy correction with $H_p$ in $n$-th order with
\begin{equation}
\begin{aligned}
\delta E^{(n)}=\sum_{\{\psi_i\}}\frac{\langle 0|H_p|\psi_1\rangle\langle \psi_1|H_p|\psi_2\rangle\cdot\cdot\cdot\langle \psi_{n-1}|H_p|0\rangle}{(E_0-E_{\psi_1})\cdot\cdot\cdot(E_0-E_{\psi_{n-1}})},
\end{aligned}
\label{perturbation}
\end{equation}
where $E_0$ is the energy for the classical configuration, $|0\rangle$, and $E_{\psi_i}$ is the energy for intermediate state denoted by $|\psi_i\rangle$. 
The summation will run over every possible intermediate processes for a given order. For any order of perturbation, its energy correction is obtained under several rules:

(1) Each perturbation is evaluated within a linked cluster to ensure the extensiveness of energy correction. On each bond of such cluster $H_p$ can be acted for more than one time and the total number of execution of $H_p$ determines the order of perturbation.

(2) Any series must start and end with $J^{(1)}_{ij}$ and $J^{(3)}_{ij}$, since we have rotated the local coordinate so that for every site, $\textbf{S}^{(0)}_i=S\hat{e}^z_i$.

(3) $J^{(3)}_{ij}$ and $B_i$ only consider one spin flip and thus there must be an even number for such process in a series. Because of this constraint, one can always find a lower order series without these two terms and therefore we can ignore them in search of leading order correction.

\begin{figure}[tbp]
\centering
 	\includegraphics[width=1.0\columnwidth]{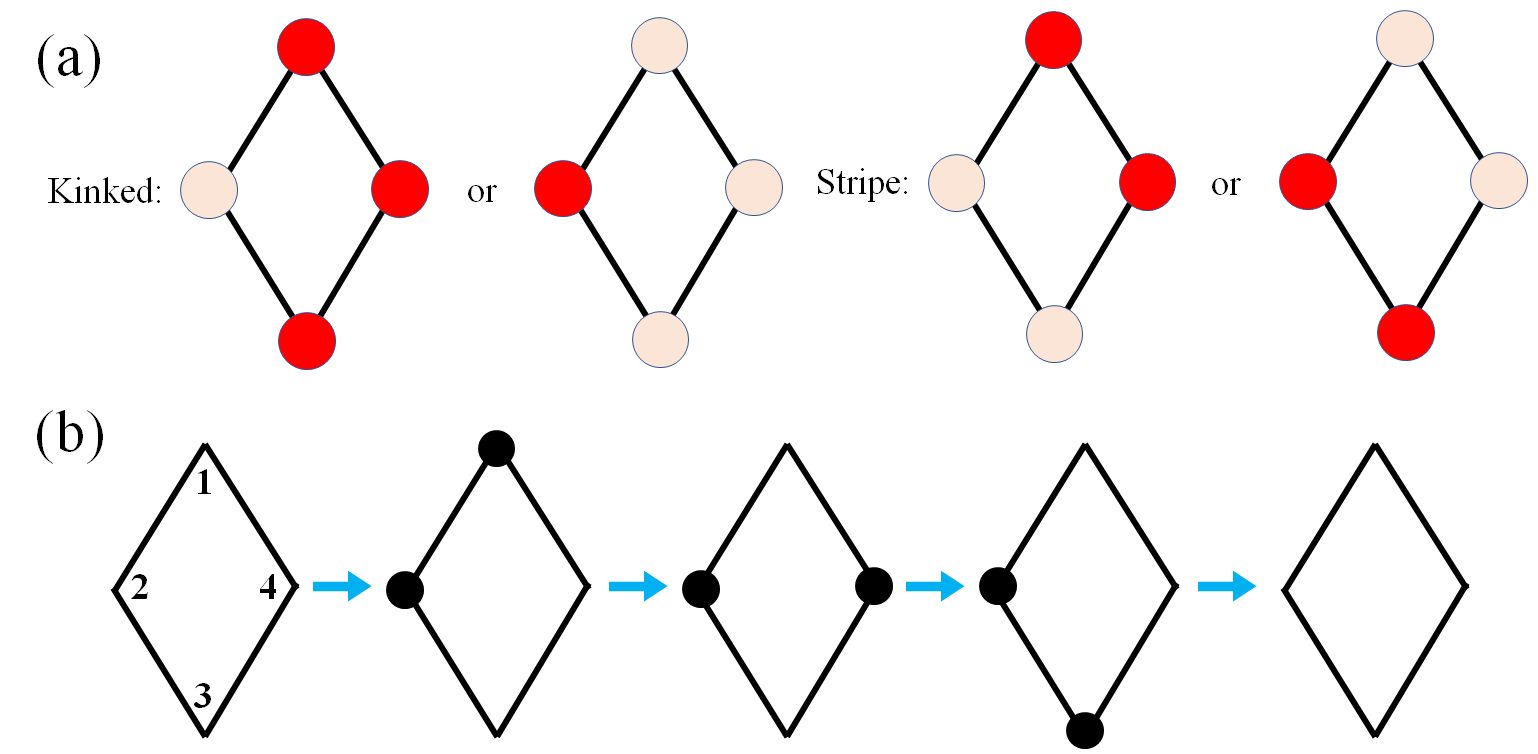}
\caption{(a) The four-site diamond cluster that distinguishes kinked and stripe states. Red and pink dots indicate sublattice $a$ and $b$ separately. For kinked state it has two configurations with three sublattice $a$ or $b$ while for stripe state the numbers of $a$ and $b$ sublattice are equal. (b) An example for the fourth-order tunneling process within the diamond cluster. The perturbative process corresponds to $J^{(1)}_{12} \rightarrow J^{(2)}_{14} \rightarrow J^{(2)}_{34} \rightarrow J^{(1)}_{23}$. Black dots indicate the places where spins have been flipped.}
\label{fig2}
\end{figure}

Because of the above rules, we know that effective leading-order perturbative series must start and end with $J^{(1)}_{ij}$ and are composed of $J^{(1)}_{ij}$ and $J^{(2)}_{ij}$. 
In Eq.~(\ref{perturbation}), the denominator considers the energy difference between classical and intermediate configurations. It is given by
\begin{equation}
\begin{aligned}
E_0-E_{\psi}=\langle 0|H|0\rangle-\langle \psi|H|\psi\rangle=-\langle \psi|H_{\text{unperturbed}}|\psi\rangle.
\end{aligned}
\label{difference}
\end{equation}

To adopt RSPT, first we rotate the local coordinates so that $\textbf{S}^{(0)}_i=S\hat{e}^z_i$. Our classical configurations are composed of two different sub-lattices representing the up and down spins
\begin{equation}
\begin{aligned}
\hat{e}^z_a=(\sin{\phi_a},0,\cos{\phi_a}),\;\hat{e}^z_b=(\sin{\phi_b},0,\cos{\phi_b}),
\end{aligned}
\label{coordz}
\end{equation}
and then we choose
\begin{equation}
\begin{aligned}
&\hat{e}^\pm_a=\frac{1}{2}(\cos{\phi_a},\pm i,-\sin{\phi_a}),\\
&\hat{e}^\pm_b=\frac{1}{2}(\cos{\phi_b},\pm i,-\sin{\phi_b}),
\end{aligned}
\label{coordx}
\end{equation}
where $a$ and $b$ denote two different sub-lattices. Without the loss of generality, we choose the planar spin orienting along $\hat{x}$ axis. 
In fact, in the classical limit we can simply assign $\phi_a=0$ for up spin and $\phi_b=\pi$ for down spin.
Since we hope that our derivation can also be useful in the more general scenarios breaking the $U(1)$ symmetry, we take the off-diagonal order into our consideration.
It is also clear to see that by adopting the same coordinate of sub-lattice $a$ or $b$ for every site in Eq.~(\ref{HamiltonianRSPT2}), we can resume the original Hamiltonian, Eq.~(\ref{HamiltonianT}), where the translation invariance is accompanied.

To estimate the leading-order correction, we need to first look for the minimal closed cluster which differentiates the stripe and kinked patterns, and it is the vertical diamonds whose longer diagonal line is parallel to $\hat{y}$ in Fig.~\ref{fig1}.
In Fig.~\ref{fig2}(a) we demonstrate the minimal different cluster for kinked and stripe states.
Then, we need to flip the spins through the intermediate process on the diamond bonds to evaluate the energy correction. Since there are four bonds on each diamond, the leading-order correction which we are looking for is of the fourth order.
One of the possible intermediate processes is shown in Fig.~\ref{fig2}(b).

The detailed equations and derivation of RSPT are shown in Appendix~\ref{AppenA} and here we only demonstrate the final results.
While the degeneracy stays intact when $J=0$, turning on the spin-flip coupling will immediately introduce an energy correction to different patterns.
For the stripe and kinked patterns, we have $\phi_a=0$ and $\phi_b=\pi$ for the classical configurations.
Under this assumption our equations can be largely simplified. More importantly, $J^{d(2)}_3$ becomes zero and thus only a few terms remain in Eqs.~(\ref{kinkedec}) and (\ref{stripeec}).
The energy correction terms then become
\begin{equation}
\begin{aligned}
&\delta E^{(4)}_\text{kinked}=\frac{2J^4}{(\Gamma_3)^2\Gamma_4},\\
&\delta E^{(4)}_\text{stripe}=\frac{2J^4}{(\Gamma_3)^2\Gamma_4}+\frac{2J^4}{(\Gamma_3)^2\Gamma_5}.
\end{aligned}
\label{ecpb}
\end{equation}
Since $J^z_h$ is repulsive and thus $\Gamma_5=-8S^2J^z_h<0$~(see Appendix~\ref{AppenA}), the stripe configuration is favored and would be selected.
This conclusion coincides with the previous one for the frustrated colloidal soft matter system, whose thermal OBD has been studied through the entropy estimation~\cite{Shokef11804}.
There, the authors have shown that the straight stripe configuration possesses lower free energy, despite the fact that during the cooling process the energy barrier among different configurations will eventually lead to a mixed pattern.
The effective Hamiltonian of such system can be written down as an antiferromagnetic Ising model with deformation in the triangular lattice, similar to our system considered here.
At last, we emphasize that since the leading-order correction is of the fourth order, the sign of $J$ does not affect the final conclusion.
In Section~\ref{CMFT}, we will consider an effective model with the ferromagnetic $J$.

\subsection{\label{iPEPS}iPEPS calculation}

\begin{table*}
\centering
\begin{tabular}{ccccccccc}\hline\hline
             & \begin{tabular}[c]{@{}c@{}}~$D=2$~\\\end{tabular} & & & & ~~$D=3$~~& &    \\ 
\hline
$J/J_z$  & $\langle H_s \rangle$  & $m$ & $\tilde{S}^z$ & $\tilde{S}^+$  & $\langle H_s \rangle$  & $m$ & $\tilde{S}^z$ & $\tilde{S}^+$ \\
\hline
~$0.1$~ & -0.26327  & 0.47966 & 0.47966  & 0.00002 & -0.26336 & 0.47880 & 0.47880  & 0.00004 \\
~$0.2$~ & -0.29746  & 0.43363 & 0.43363  & 0.00004 & -0.29843 & 0.42404 & 0.42404  & 0.00007 \\
~$0.3$~ & -0.34539  & 0.38197 & 0.38197  & 0.00004 & -0.34853 & 0.35545 & 0.35545  & 0.00032 \\
~$0.4$~ & -0.40150  & 0.33782 & 0.33782  & 0.00007 & -0.40753 & 0.29673 & 0.29673  & 0.00108 \\
~$0.5$~ & -0.46177  & 0.30808 & 0.30808  & 0.00001 & -0.47140 & 0.25309 & 0.25308  & 0.00174 \\
~$0.6$~ & -0.52501  & 0.28392 & 0.28392  & 0.00003 & -0.53779 & 0.22126 & 0.22122  & 0.00372 \\
~$0.7$~ & -0.58987  & 0.26586 & 0.26586  & 0.00005 & -0.60623 & 0.18589 & 0.18586  & 0.00359 \\
\hline\hline
\end{tabular}
\caption{We provide the values of energies and orders for several $J/J_z$, with bond dimension $D=2$ or 3.}
\label{tab1}
\end{table*}

Since RSPT applies well only for small $J$, we would like to see if the conclusion still holds when the quantum fluctuation is strong.
For this purpose, we adopt the two-dimensional~(2D) tensor network ansatz, the infinite projected entangled-pair state~(iPEPS), to numerically calculate the simplified Eq.~(\ref{HamiltonianT})
\begin{equation}
\begin{aligned}
H_s&=J\sum_{\langle ij \rangle}(\hat{S}^+_{i}\hat{S}^-_{j}+\hat{S}^-_i\hat{S}^+_j)+J^z\sum_{\langle ij\rangle\in h} \hat{S}^z_i\hat{S}^z_j,
\end{aligned}
\label{HamiltonianSS}
\end{equation}
where we consider the extreme case and ignore both the Ising coupling along the diagonal direction and the longitudinal field.
By properly choosing the pre-designated unit-cell size~(in this work $2\times2$ or $4\times2$ as shown in Fig.~\ref{fig1}) and optimizing the $d\times D\times D\times D\times D$ tensors, where $d$ represents the dimension of local Hilbert space and $D$ is the virtual bond dimension, this tensor network ansatz serves as a good variational wavefunction for the quantum many-body systems.
We provide some details of iPEPS in Appendix~\ref{AppenB}.

To numerically estimate the ground state ansatz of Eq.~(\ref{HamiltonianSS}), we need to adopt numerous trials starting from different initial setups.
It is because we already recognize that there are many competing states~(stripe, kinked, or mixed states) so that we need to avoid the simulation being trapped in some undesirable local minima.
As a result, besides the random initial tensors we also start our calculation by constructing the distinct product states, corresponding to the stripe and kinked configurations separately.
Since the product state is of $D=1$, we then enlarge our tensors to the assigned $D$, with extra tensor elements being small random numbers.
Through this measure, we guide the optimization toward obtaining the stripe or kinked state.
After the convergence is reached, the averaged magnetization is defined as
\begin{equation}
\begin{aligned}
m=\frac{1}{N_C}\sum_{i\in C}\sqrt{\langle \hat{S}^x_i\rangle^2+\langle \hat{S}^y_i\rangle^2+\langle \hat{S}^z_i\rangle^2} ,
\end{aligned}
\label{m_avg}
\end{equation}
where $C$ means the repeating unit cell and $N_C$ is the number of lattice sites within the cell.
For the stripe pattern since there are two possible choices for the iPEPS unit cell and they correspond to the wave numbers $(0,\pi)$ and $(\pi,\pi)$ after Fourier transformation, we thus define the two related orders for the stripe pattern. The first one is the diagonal magnetic moment
\begin{equation}
\begin{aligned}
\tilde{S}^z=\frac{1}{N_C}\sum_{i\in C}\sum_{\textbf{k}\in\textbf{k}_1,\textbf{k}_2}\langle \hat{S}^z_i\rangle e^{i\textbf{k}\cdot \textbf{r}_i},
\end{aligned}
\label{Szorder}
\end{equation}
where $\textbf{k}_1=(\pi,\pi)$ and $\textbf{k}_2=(0,\pi)$. $\textbf{r}_i$ represents the coordinate $(r_i^x,r_i^y)$ for each site within the unit cell. For the off-diagonal order, we define
\begin{equation}
\begin{aligned}
\tilde{S}^+=\frac{1}{N_C}\sum_{i\in C}|\langle \hat{S}^+_i\rangle|.
\end{aligned}
\label{Sporder}
\end{equation}

In Table~\ref{tab1} we provide the lowest energies as well as the orders for several different $J$.  
We can clearly see that for each $J/J_z$ we have $m\approx \tilde{S}^z$, meaning that despite the different initial trials the states of lowest energy always belong to the stripe state.
Although the off-diagonal orders are almost zero, $\langle\hat{S}^z\rangle$ is highly supressed due to the quantum effect, suggesting that the initial setup for RSPT might not be well applicable.
However, our simulation indicates that the conclusion does not change and the stripe state keeps being the ground state, selected by the quantum OBD effect. 
It is also worth mentioning that for $D=3$ and $J>0.3$ no matter what initial states we adopt the ground state ansatz always evolves toward the stripe state. 
We then conclude that our system, where the circular Rydberg atoms are placed in a deformed triangular optical lattice, favors the existence of stripe state.

\section{\label{CMFT}Dipolar hard-core bosons}

In the previous section we have revealed the quantum OBD effect for the targeted Rydberg system and identified that the stripe state is more stable.
Next, we aim to unveil the possible underlying phases for the artificial Hamiltonian of dipolar bosons in the triangular optical lattice, which can be generated with cold-atom simulators.
The realization of placing magnetic cold atoms in optical lattices has been previously reported for erbium~\cite{doi:10.1126/science.aac9812} and ytterbium~\cite{TAKAHASHI2022PJA9804B-01}, where a three-dimensional~(3D) optical lattice is generated with two horizontal lasers in the x-y plane and one vertical laser reflecting the direction of gravity. 
By confining the atoms in a 3D lattice it would help elongate the lifetime by preventing the inelastic collision of atoms~\cite{PhysRevLett.108.080405, doi:10.1126/science.aac6400}.
Since the vertical tunneling can be reduced by adopting a laser with a longer wavelength, and thus a quasi-2D system can be synthesized.
Ref.~\cite{doi:10.1126/science.aac9812} provides a good example using the laser with wavelength $(\lambda_x,\lambda_y,\lambda_z)=(532,532,1064)$ in the unit of nanometer, creating the effective 3D optical lattice with size $(l_x,l_y,l_z)=(266,266,532)$ and a trapping potential $V(x,y,z)=V_xcos^2(k_xx)+V_ycos^2(k_yy)+V_zcos^2(k_zz)$ where $V_i$ is the lattice depth and $k_i$ is the wavevector in different direction.
After preparing the Bose-Einstein condensate~(BEC) of the target atoms through the optical dipole trap~(ODT), the BEC cloud is then adiabatically loaded to the optical lattice within a time scale of millisecond.
In the end we obtain the long-lived system whose lifetime scales up to the unit of second~\cite{PhysRevLett.108.080405, doi:10.1126/science.aac9812}.
For our system, however, it demands a triangular optical lattice for its realization and Ref.~\onlinecite{Becker_2010} has unveiled this possibility using Rb BEC.
There, they adopt a three-laser trapping within the x-y plane whose wavevectors mutually share a 120$^\circ$ enclosing angle.
Optionally, the recent proposal of using optical tweezers to trap the cold atoms might also provide a more flexible platform in designing the quantum simulators~\cite{PhysRevX.9.021039, doi:10.1126/science.aax1265}.

\begin{figure}[tbp]
\centering
 	\includegraphics[width=1.0\columnwidth]{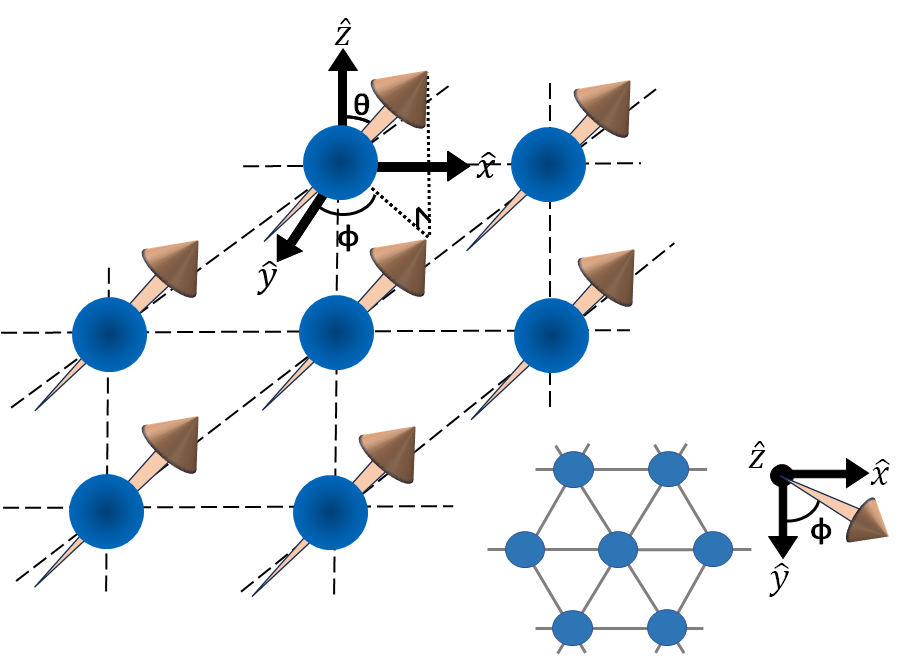}
\caption{Schematic demonstration of dipolar interaction in the triangular lattice. Dots denote the lattice site and arrows represent the dipole polarization. Such polarization can be parametrized with polar ($\theta$) and azimuthal ($\phi$) angles. The lower right panel shows the projection of lattice from above for a better demonstration of the azimuthal angle.}
\label{fig3}
\end{figure}

Next we head to discuss the Hamiltonian of interest and its realization through cold-atom simulators.
The magnetic quantum gases is composed of atoms which can be seen as magnetic dipoles even at zero magnetic field.
The origin of dipole moment comes from the spin and orbital angular momentum of electrons, as well as some minor contribution from the nuclear spin.
This fact results in the high susceptibility of atoms to an external Zeeman field, which we could use to control the overall orientation, expressed by the polar and azimuthal angles shown in Fig.~\ref{fig3}, of the magnetic dipoles~\cite{Chomaz_2023}.
This simplifies the general form of dipole-dipole interaction in Ref.~\onlinecite{Lahaye_2009} and makes it the following form
\begin{equation}
\begin{aligned}
V_{ij}=\frac{V}{r_{ij}^3}(1-3\text{cos}^2\alpha_{ij}),
\end{aligned}
\label{dipolar}
\end{equation}
where $V$ is the interactive strength and $r_{ij}=|\vec{r}_i-\vec{r}_j|$. $\alpha_{ij}$ is the included angle by the dipole moment and $\vec{r}_{ij}$.
This interaction contributes to the on-site repulsive interaction, inter-site interaction, and the density induced tunneling.
In this work we consider the hardcore limit meaning that only one atom is allowed to occupy one site in every snapshot.
This condition can be naturally fulfilled once the on-site interaction is much larger than the inter-site interaction, which can be tuned through controlling the lattice spacing~\cite{10.21468/SciPostPhys.14.5.136}.
Once the doubly occupancy is prohibited, the density induced tunneling process also disappears.
Along with the normal tunneling process, we end up obtaining the following extended Bose-Hubbard Hamiltonian with the dipolar interaction
\begin{equation}
\begin{aligned}
H=-t\sum_{\langle ij \rangle}(\hat{b}^\dagger_{i}\hat{b}_{j}+H.C.)+\sum_{\langle ij \rangle}V_{nn} \hat{n}_i \hat{n}_j-\mu \sum_i \hat{n}_i,
\end{aligned}
\label{Hamiltonian}
\end{equation}
where $\hat{b}^\dagger_{i}$ and $\hat{b}_{i}$ represent the creation and annihilation operators of the hard-core boson, with the number operator being $\hat{n}_i=\hat{b}^\dagger_{i}\hat{b}_{i}$. $V_{nn}$ denotes the $nn$ dipole-dipole interaction among bosons and we neglect its long-range tail since it decays rapidly.
We emphasize that the dipole-dipole interaction in this section is different from that for the Rydberg atoms.
As explained in Section~\ref{Rydberg}, for the circular Rydberg atoms one makes use of the electric dipole-dipole interaction of neutral atoms to generate the spin exchange coupling.
On the other hand, in this section the cold atom itself can be seen as a magnetic dipole and thus mutual interaction with the form indicated in Eq.~(\ref{dipolar}) is cast among atoms.
Moreover, the atoms act as quanta and thus tunneling effect takes place for minimizing the energy, leading to the hopping term in Eq.~(\ref{Hamiltonian}) that lowers the energy.

The interaction in Eq.~(\ref{dipolar}) can be easily tilted with an external magnetic field.
In this work we study the case when the dipole moments are tilted as $\phi=0$~(see Fig.~\ref{fig3}). Thus, for the $nn$ interaction, the two interactive terms along diagonal direction are equal and become attractive when polar angle $\theta$ is large enough. 
Therefore, there are two different terms for $V_{nn}$
\begin{equation}
\begin{aligned}
&V_{nn}^h=V\\
&V_{nn}^d=V(1-\frac{9}{4}\text{sin}^2\theta),
\end{aligned}
\label{NNInteraction}
\end{equation}
where indices $h$ and $d$ indicate the interaction along the horizontal and diagonal directions, as shown in Fig.~\ref{fig1} labeled by green and black bonds, separately. 
As a result, Eq.~(\ref{Hamiltonian}) is akin to Eq.~(\ref{HamiltonianT}) with negative $J$ since there is a one-to-one mapping by simply replacing $\hat{b}^\dagger_i \rightarrow \hat{S}^+_i$, $\hat{b}_i \rightarrow \hat{S}^-_i$, and $\hat{n}_i \rightarrow \hat{S}^z_i+\frac{1}{2}$~\cite{JPhysCondensMatter.32.455401}.
Therefore, by numerically studying Eq.~(\ref{Hamiltonian}) with different tilting angles the resulting phase diagrams cover a wide range of \textsl{XXZ}-like model with anisotropy in the triangular lattice.

\begin{figure*}[tbp]
\centering
 	\includegraphics[width=2.0\columnwidth]{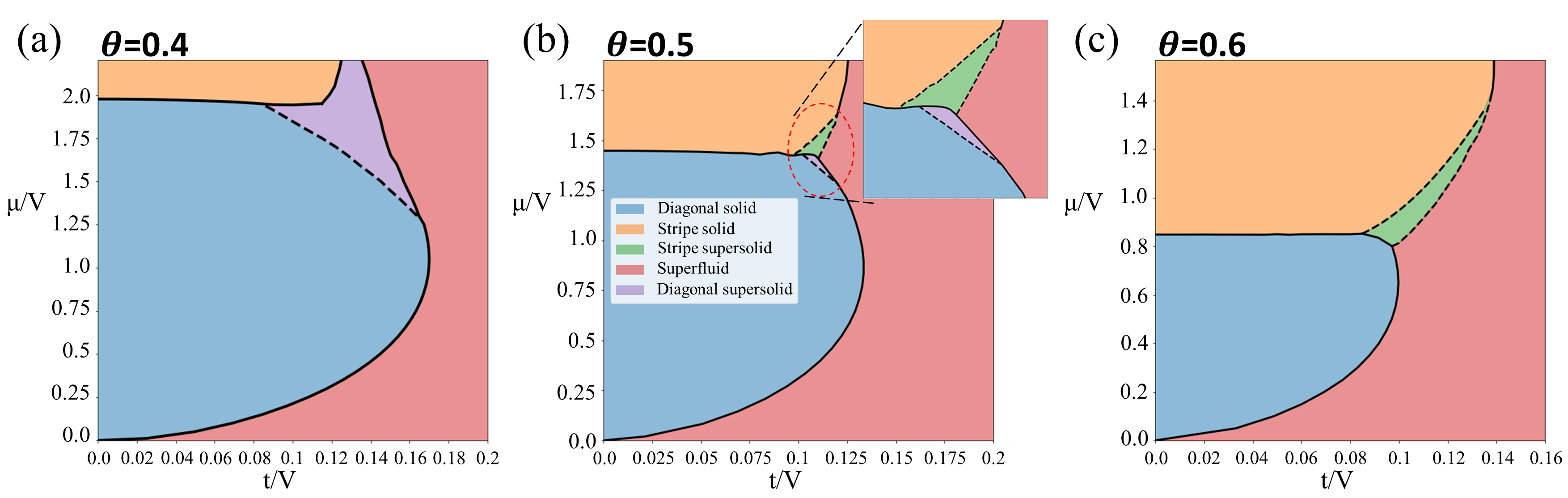}
\caption{CMFT phase diagrams with $nn$ dipolar interaction with polar angle equal to (a) 0.4, (b) 0.5, and (c) 0.6, which approximate to 0.127$\pi$, 0.159$\pi$, and 0.191$\pi$. The inset in (b) enlarges the area for two supersolid phases. Solid (dashed) phase boundaries indicate the first- (second-) order phase transition.}
\label{fig4}
\end{figure*}

In this work, we adopt CMFT~\cite{PhysRevA.85.021601, PhysRevLett.112.127203, PhysRevA.102.053306} for the construction of phase diagram.
Some Details of applying CMFT are indicated in Appendix~\ref{AppenC}.
When the strength of hopping term is small, the ground state is an ordered state with $U(1)$ symmetry preserved.
As we increase $t$, the solid order will dissolve and phase transits to the superfluid state through a first-order phase transition.
However, in some small windows between solid and superfluid phases, two orders, namely the structural and superfluid orders, can co-exist and a supersolid phase manifests.
To identify distinct phases we first introduce the order parameters adopted here.
For solid and supersolid phases, they both possess structural order which is defined as
\begin{equation}
\begin{aligned}
\tilde{n}(\textbf{k})=\frac{1}{N_C}\sum_{i\in C}\langle \hat{n}_i\rangle e^{i\textbf{k}\cdot \textbf{r}_i},
\end{aligned}
\label{struorder}
\end{equation}
where $C$ again represents the repeating unit cell whose size is equal to $3\times3$ or $4\times4$ in our CMFT calculation, and thus $N_C=9$ or 16.
As for the superfluid order, we calculate its condensate density defined as
\begin{equation}
\begin{aligned}
\rho_0=\frac{1}{N_C}\sum_{i\in C}|\langle \hat{b}_i\rangle|^2.
\end{aligned}
\label{condens}
\end{equation}
The above two order parameters help identify different phases in our numerical phase diagrams.

We then conduct the CMFT calculation and plot the phase diagrams for several polar angles in Fig.~\ref{fig4}. 
When tilting is small~(Fig.~\ref{fig4}(a)), the diagonal solid~(blue) and supersolid~(purple) dominate the phase diagram before entering the superfluid phase. Its structural factor has the modulating momentum $(2\pi/3,2\pi/3)$, and is named after the diagonal stripe in our earlier work~\cite{PhysRevA.102.053306}. 
In the supersolid phase the $\tilde{n}(2\pi/3,2\pi/3)$ and $\rho_0$ orders co-exist. This phase appears after a continuous transition from the diagonal solid. Further enlarging $t$, solid order disappears and the phase transits into superfluid~(red) phase discontinuously.

On the other hand, the central orange lobe indicates the classically highly degenerate solid configurations~(stripe, kinked, and mixed).
Thanks to the diagnosis of RSPT and iPEPS in the previous section, we have learned that the true ground state must be in the stripe configuration, although the numerical energies of stripe and kinked configurations by CMFT are very close to each other due to the small $t$.
The green phase in Fig.~\ref{fig4}(b) and \ref{fig4}(c) represents a supersolid phase coming from the stripe solid and thus we call it the stripe supersolid.
Note that our results indicate that the transition from stripe solid to stripe supersolid, as well as the one from stripe supersolid to superfluid are both continuous.

One can see from Fig.~\ref{fig4} that as we further tilt the polar angle, the central lobe grows and its corresponding supersolid phase also starts to appear~(Fig.~\ref{fig4}(b)). 
Finally, as the polar angle is large enough, dominant ordered phases in the phase diagram will be replaced by the stripe solid and supersolid~(Fig.~\ref{fig4}(c)). 
The scenario described here is just opposite to the one in our previous work, where a square lattice effectively shifts to a triangular one by tilting the polar angle~\cite{PhysRevA.102.053306}.
Here, we start from the triangular lattice where the corresponding phase possesses three sub-lattices~(see, e.g., Fig.~3 in Ref.~\cite{PhysRevA.102.053306} for explanation), and eventually ends up with a square one with only two sub-lattices, according to the phase diagram.
At last, we emphasize that including the long-range tail of dipole-dipole interaction will automatically lift the degeneracy even in the classical limit.
In such scenario we expect a even more fruitful phase diagram with many different ground-state configurations but it requires a highly precise experimental tool to sort out.
We will leave the further studies for this issue in the future work.

\section{\label{conclu}Conclusion}

In this work we study two different systems of the dipolar quantum simulator.
Placing the circular Rydberg atoms in a deformed triangular optical lattice, a sub-extensive degeneracy can be realized and the spin exchange interaction, arising from the electric dipole-dipole interaction of Rydberg atoms, leads to the quantum OBD effect. 
The degeneracy of lowest-energy configuration in the classical limit is caused by the deformation, making the van der Waals repulsion only be seen in the horizontal direction between nearby sites. 
Once the spin exchange term is introduced, the overall Hamiltonian is akin to the \textsl{XXZ} model with anisotropic interactive potential.
With RSPT and iPEPS, we predict that the stripe configuration is the true ground state, in coincidence with the thermal counterpart of this OBD effect according to the previous work~\cite{Shokef11804}.
Next, we consider an extended Bose-Hubbard model with magnetic dipole-dipole interaction in the optical lattice, which is related to the cold-atom quantum simulator.
We then provide the phase diagrams for the $nn$ dipolar hard-core bosonic Hamiltonian with different tilting angles in the triangular lattice, using CMFT.
We exploit the competing scenario proposed in our earlier work~\cite{PhysRevA.102.053306} and demonstrate that the effective lattice structure shifts from triangular lattice to a square one, as well as various different phases including the supersolid.

The technique of neutral atoms for quantum computing has recently attracted huge attention. 
By placing the Rydberg atoms with the desirable geometric controls, large-scale quantum Hamiltonian can be simulated with the quantum analog simulators~\footnote{See \href{https://github.com/QuEraComputing/quera-education}{https://github.com/QuEraComputing/quera-education} for some educational materials.}.
However, the current platforms mainly focus on the Ising model and its variants. In this work, we push one step forward and combine the spin exchange interaction as well as the van der Waal repulsion. 
We propose a possible experimental setup using the circular Rydberg atoms in optical lattice for realization, which benefits a further investigation by experimental groups.
Our analytical and numerical studies, as well as the conclusion, might be useful benchmarks once the simulator can be put into practice, instead of checking the full phase diagram~(see, e.g., Section III.B of Ref.~\onlinecite{PhysRevX.8.011032}).
Moreover, controlling the anisotropy of the triangular lattice, we might reach a novel regime where the order disappears and a state with larger entanglement entropy manifests~\cite{PNAS2021}.
This would benefit further studies with quantum simulators as well as numerical calculation through quantum Monte Carlo or tensor networks, and we will leave it for the future consideration.
It is also important to note that our proposal unveils a possible protocol in studying the dynamics of glassy phases through quantum simulator.
In Ref.~\cite{Shokef11804} the authors revealed that although the straight stripe benefits from lower free energy and should be more favorable in arbitrarily low temperature region, the free-energy barrier between different configurations results in a metastable disorder state while cooling down the temperature.
Similarly, if we introduce the spin exchange coupling in the same manner a non-trivial glassy dynamics~\cite{PhysRevX.7.021030} can likely be simulated.

Moreover, the quantum simulator made of cold atoms with magnetic dipoles plays another important role in simulating the extended Bose-Hubbard model.
While each atom can be seen as a magnetic dipole, their collective alignment can be manipulated by an external field.
By tilting the polar angle of dipoles, the corresponding phase diagram changes and peculiar phases also appear.
This implies that through some simple action such as tilting the polar angle, many extraordinary physical scenario can be artificially realized through such quantum simulators.
In sum, we believe that our both proposals indicate alternative paths in applying quantum simulators for studying the intriguing physics and demonstrate their great potential for artificial many-body systems.

\section{\label{acknow}Acknowledgement}
W.-L.T. acknowledge the useful discussion with Mike Zhitomirsky during the International Conference on Strongly Correlated Electron Systems 2023~(SCES 2023), Pedro Lopes from QuEra Computing Inc., Thomas Ayral during the SQAI-NCTS Workshop on Tensor Network and Quantum Embedding in Tokyo, Hyun-Yong Lee, Tsuyoshi Okubo, S. R. Ghazanfari, and Xinliang Lyu. 
Authors also acknowledge the Advanced Study Group (ASG) Tensor Network Approaches to Many-Body Systems organized by the Center for Theoretical Physics of Complex Systems (PCS) of the Institute for Basic Science (IBS) in Daejeon, Korea.
H.-K.W. is supported by JQI-NSF-PFC (supported by NSF grant PHY-1607611).
N.K. is supported by JSPS KAKENHI Grants No. JP19H01809 and No. JP23H01092.
T.S. is supported by JSPS KAKENHI Grant. No. 21K03390.
W.-L.T. is supported by the Center of Innovations for Sustainable Quantum AI~(JST Grant Number JPMJPF2221).
This work is also mainly supported by the above-mentioned JST Grant.

\begin{figure*}[tbp]
\centering
 	\includegraphics[width=2.05\columnwidth]{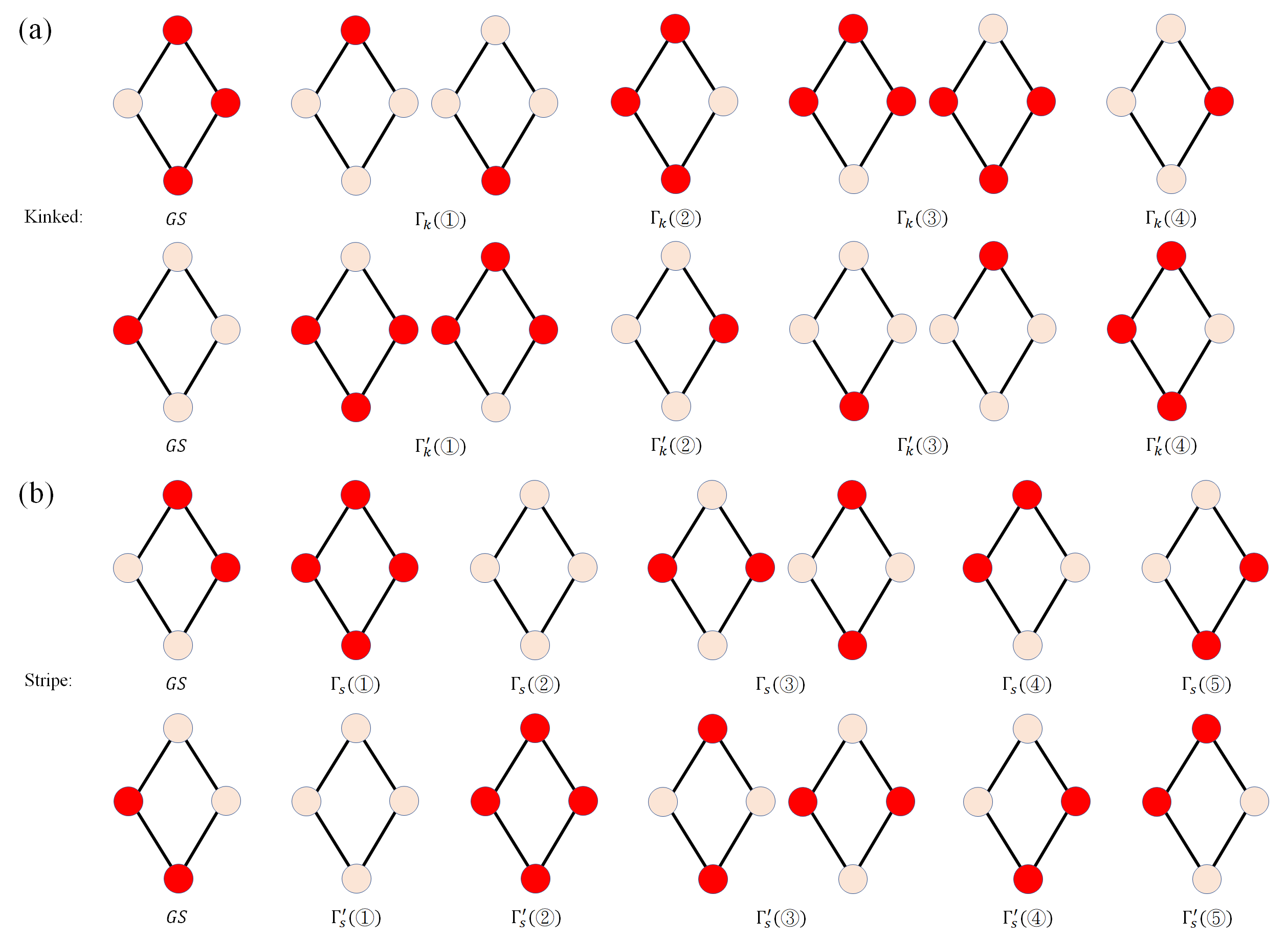}
\caption{Possible configurations during the perturbation process for (a)kinked and (b)stripe states. For the ground state there are two configurations in kinked state and thus their evolving must be considered separately. Different configurations that give equal denominator contribution are labeled with the same $\Gamma$.}
\label{fig5}
\end{figure*}

\appendix

\section{\label{AppenA}Details for the real space perturbation theory}

\begin{table*}
\centering
\begin{tabular}{ccccccccc}\hline\hline

Kinked & & \\
\hline
~$\Gamma_k(\circled{1})$~ & $4S^2\Gamma_a+8S^2(J^z_d+J^z_h)\cos\phi_a\cos\phi_b+32S^2J\sin\phi_a\sin\phi_b-2E_\textbf{B}^a$  &  \\
~$\Gamma_k(\circled{2})$~ & $4S^2(\Gamma_a+\Gamma_b)+4S^2(2J^z_d+J^z_h)\cos\phi_a\cos\phi_b+24S^2J\sin\phi_a\sin\phi_b-E_\textbf{B}^a-E_\textbf{B}^b$  &   \\
~$\Gamma_k(\circled{3})$~ & $4S^2(\Gamma_a+\Gamma_b)+4S^2(J^z_d+2J^z_h)\cos\phi_a\cos\phi_b+24S^2J\sin\phi_a\sin\phi_b-E_\textbf{B}^a-E_\textbf{B}^b$  &   \\
~$\Gamma_k(\circled{4})$~ & $8S^2\Gamma_a+8S^2(J^z_d+J^z_h)\cos\phi_a\cos\phi_b+32S^2J\sin\phi_a\sin\phi_b-2E_\textbf{B}^a$  &   \\
~$\Gamma'_k(\circled{1})$~ & $4S^2\Gamma_b+8S^2(J^z_d+J^z_h)\cos\phi_a\cos\phi_b+32S^2J\sin\phi_a\sin\phi_b-2E_\textbf{B}^b$  &   \\
~$\Gamma'_k(\circled{4})$~ & $8S^2\Gamma_b+8S^2(J^z_d+J^z_h)\cos\phi_a\cos\phi_b+32S^2J\sin\phi_a\sin\phi_b-2E_\textbf{B}^b$  &   \\
\hline\hline
Stripe  &  & \\
\hline
~$\Gamma_s(\circled{1})$~ & $4S^2\Gamma_b+8S^2(J^z_d+J^z_h)\cos\phi_a\cos\phi_b+32S^2J\sin\phi_a\sin\phi_b-2E_\textbf{B}^b$  &  \\
~$\Gamma_s(\circled{2})$~ & $4S^2\Gamma_a+8S^2(J^z_d+J^z_h)\cos\phi_a\cos\phi_b+32S^2J\sin\phi_a\sin\phi_b-2E_\textbf{B}^a$  &  \\
~$\Gamma_s(\circled{3})$~ & $4S^2(\Gamma_a+\Gamma_b)+4S^2(J^z_d+2J^z_h)\cos\phi_a\cos\phi_b+24S^2J\sin\phi_a\sin\phi_b-E_\textbf{B}^a-E_\textbf{B}^b$  &   \\
~$\Gamma_s(\circled{4})$~ & $4S^2(\Gamma_a+\Gamma_b)+4S^2(2J^z_d+J^z_h)\cos\phi_a\cos\phi_b+24S^2J\sin\phi_a\sin\phi_b-E_\textbf{B}^a-E_\textbf{B}^b$  &   \\
~$\Gamma_s(\circled{5})$~ & $4S^2(\Gamma_a+\Gamma_b)+8S^2(J^z_d+J^z_h)\cos\phi_a\cos\phi_b+32S^2J\sin\phi_a\sin\phi_b-E_\textbf{B}^a-E_\textbf{B}^b$  &   \\
\hline\hline
\end{tabular}
\caption{Different $\Gamma$ for the kinked configuration. $E_\textbf{B}^a=2S\textbf{B}\cdot \hat{e}^z_a$ and $E_\textbf{B}^b=2S\textbf{B}\cdot \hat{e}^z_b$. $\Gamma_a=J^z_d\cos^2\phi_a+2J\sin^2\phi_a$ and $\Gamma_b=J^z_d\cos^2\phi_b+2J\sin^2\phi_b$.}
\label{tab2}
\end{table*}

In this Appendix, we work on the details for RSPT.
With the choice for the local coordinate shown in the main text~(Eqs.~(\ref{coordz}) and (\ref{coordx})), we can now write down the corresponding $J^{(1)}_{ij}$ and $J^{(2)}_{ij}$ terms. For diagonal bonds we have
\begin{equation}
\begin{aligned}
&J^{d(1)}_{aa}=\frac{1}{4}\sin^2\phi_a(J^z_{d}-2J)S^+_aS^+_a+H.C.\\
&J^{d(1)}_{bb}=\frac{1}{4}\sin^2\phi_b(J^z_{d}-2J)S^+_bS^+_b+H.C.\\
&J^{d(1)}_{ab}=\frac{1}{4}[2J(\cos\phi_a\cos\phi_b-1)+J^z_{d}\sin\phi_a\sin\phi_b]S^+_aS^+_b\\
&\;\;\;\;\;\;\;\;\;\;\;\;\;+H.C.,
\end{aligned}
\label{v1}
\end{equation}
where we assign $J^{d(1)}_{1}=\frac{1}{4}\sin^2\phi_a(J^z_{d}-2J)$, $J^{d(1)}_{2}=\frac{1}{4}\sin^2\phi_b(J^z_{d}-2J)$, and $J^{d(1)}_{3}=\frac{1}{4}[2J(\cos\phi_a\cos\phi_b-1)+J^z_{d}\sin\phi_a\sin\phi_b]$. And
\begin{equation}
\begin{aligned}
&J^{d(2)}_{aa}=\frac{1}{4}[2J(1+\cos^2\phi_a)+J^z_{d}\sin^2\phi_a]S^+_aS^-_a+H.C.\\
&J^{d(2)}_{bb}=\frac{1}{4}[2J(1+\cos^2\phi_b)+J^z_{d}\sin^2\phi_b]S^+_bS^-_b+H.C.\\
&J^{d(2)}_{ab}=\frac{1}{4}[2J(1+\cos\phi_a\cos\phi_b)+J^z_{d}\sin\phi_a\sin\phi_b]S^+_aS^-_b\\
&\;\;\;\;\;\;\;\;\;\;\;\;\;+H.C.,
\end{aligned}
\label{v2}
\end{equation}
where we again set $J^{d(2)}_{1}=\frac{1}{4}[2J(1+\cos^2\phi_a)+J^z_{d}\sin^2\phi_a]$, $J^{d(2)}_{2}=\frac{1}{4}[2J(1+\cos^2\phi_b)+J^z_{d}\sin^2\phi_b]$, and $J^{d(2)}_{3}=\frac{1}{4}[2J(1+\cos\phi_a\cos\phi_b)+J^z_{d}\sin\phi_a\sin\phi_b]$.
For the horizontal bonds we simply need to replace $J^z_{d}$ in Eq.~(\ref{v1}) and (\ref{v2}) with $J^z_{h}$, but in fact it does not affect the leading-order correction since our cluster of interest is the vertical diamond~(Fig.~\ref{fig2}(a)).
Moreover, all possible intermediate configurations as well as their corresponding $-\langle \psi|H_{\text{unperturbed}}|\psi\rangle$, labeled by different $\Gamma$ and composing the denominator of energy correction, also need to be considered.
For this purpose, we first list all the possible intermediate configurations during the spin flipping process, shown in Fig.~\ref{fig5}, with each configuration's corresponding minus unperturbed energy, $E_0-E_{\psi}$, labeled by different $\Gamma$.
We list their formula in Table~\ref{tab2}. Note that $\Gamma'_k(\circled{2})=\Gamma_k(\circled{2})$, $\Gamma'_k(\circled{3})=\Gamma_k(\circled{3})$, and $\Gamma'_s(\circled{1})=\Gamma_s(\circled{2})$, $\Gamma'_s(\circled{2})=\Gamma_s(\circled{1})$, $\Gamma'_s(\circled{3})=\Gamma_s(\circled{3})$, $\Gamma'_s(\circled{4})=\Gamma_s(\circled{4})$, $\Gamma'_s(\circled{5})=\Gamma_s(\circled{5})$.
For simplicity, we further conclude that
\begin{equation}
\begin{aligned}
&\Gamma_1=\Gamma_k(\circled{1})=\Gamma_s(\circled{2})=\Gamma'_s(\circled{1}),\\
&\Gamma_2=\Gamma'_k(\circled{1})=\Gamma_s(\circled{1})=\Gamma'_s(\circled{2}),\\
&\Gamma_3=\Gamma_k(\circled{3})=\Gamma'_k(\circled{3})=\Gamma_s(\circled{3})=\Gamma'_s(\circled{3}),\\
&\Gamma_4=\Gamma_k(\circled{2})=\Gamma'_k(\circled{2})=\Gamma_s(\circled{4})=\Gamma'_s(\circled{4}),\\
&\Gamma_5=\Gamma_s(\circled{5})=\Gamma'_s(\circled{5}).
\end{aligned}
\label{gammaf}
\end{equation}
Along with $\Gamma_k(\circled{4})$ and $\Gamma'_k(\circled{4})$, we have obtained all the ingredients needed for the denominators of the energy corrections.

Finally, through considering all possible intermediate processes akin to the one in Fig.~\ref{fig2}(b), we obtain the energy correction. For the kinked state,
\begin{equation}
\begin{aligned}
&\delta E^{(4)}_\text{kinked}=\frac{(J^{d(1)}_1)^2(J^{d(2)}_3)^2}{(\Gamma_1)^2\Gamma_4}+\frac{(J^{d(1)}_2)^2(J^{d(2)}_3)^2}{(\Gamma_2)^2\Gamma_4}\\
&+\frac{(J^{d(1)}_3)^2(J^{d(2)}_1)^2}{(\Gamma_3)^2\Gamma_4}+\frac{(J^{d(1)}_3)^2(J^{d(2)}_2)^2}{(\Gamma_3)^2\Gamma_4}\\
&+\frac{4J^{d(1)}_1J^{d(1)}_3J^{d(2)}_1J^{d(2)}_3}{\Gamma_1\Gamma_3\Gamma_k(\circled{4})}+\frac{4J^{d(1)}_2J^{d(1)}_3J^{d(2)}_2J^{d(2)}_3}{\Gamma_2\Gamma_3\Gamma'_k(\circled{4})}\\
&+\frac{2J^{d(1)}_1J^{d(1)}_3J^{d(2)}_1J^{d(2)}_3}{\Gamma_1\Gamma_3\Gamma_4}+\frac{2J^{d(1)}_2J^{d(1)}_3J^{d(2)}_2J^{d(2)}_3}{\Gamma_2\Gamma_3\Gamma_4},
\end{aligned}
\label{kinkedec}
\end{equation}
and for stripe state,
\begin{equation}
\begin{aligned}
&\delta E^{(4)}_\text{stripe}=\frac{2(J^{d(1)}_3)^2J^{d(2)}_1J^{d(2)}_2}{(\Gamma_3)^2\Gamma_4}+\frac{2J^{d(1)}_1J^{d(1)}_2(J^{d(2)}_3)^2}{\Gamma_1\Gamma_2\Gamma_4}\\
&+\frac{2(J^{d(1)}_3)^2J^{d(2)}_1J^{d(2)}_2}{(\Gamma_3)^2\Gamma_5}+\frac{2J^{d(1)}_1J^{d(1)}_2(J^{d(2)}_3)^2}{\Gamma_1\Gamma_2\Gamma_5}\\
&+\frac{2J^{d(1)}_1J^{d(1)}_3J^{d(2)}_2J^{d(2)}_3}{\Gamma_1\Gamma_3\Gamma_4}+\frac{2J^{d(1)}_2J^{d(1)}_3J^{d(2)}_1J^{d(2)}_3}{\Gamma_2\Gamma_3\Gamma_5}\\
&+\frac{2J^{d(1)}_1J^{d(1)}_3J^{d(2)}_2J^{d(2)}_3}{\Gamma_1\Gamma_3\Gamma_5}+\frac{2J^{d(1)}_2J^{d(1)}_3J^{d(2)}_1J^{d(2)}_3}{\Gamma_2\Gamma_3\Gamma_4},
\end{aligned}
\label{stripeec}
\end{equation}
where $\delta E^{(4)}$ is the energy correction per site.
Eqs.~(\ref{kinkedec}) and (\ref{stripeec}) will then be used for studying the degeneracy lifting.

\section{\label{AppenB}Infinite projected entangled-pair state}

\begin{figure}[tbp]
\centering
 	\includegraphics[width=1.0\columnwidth]{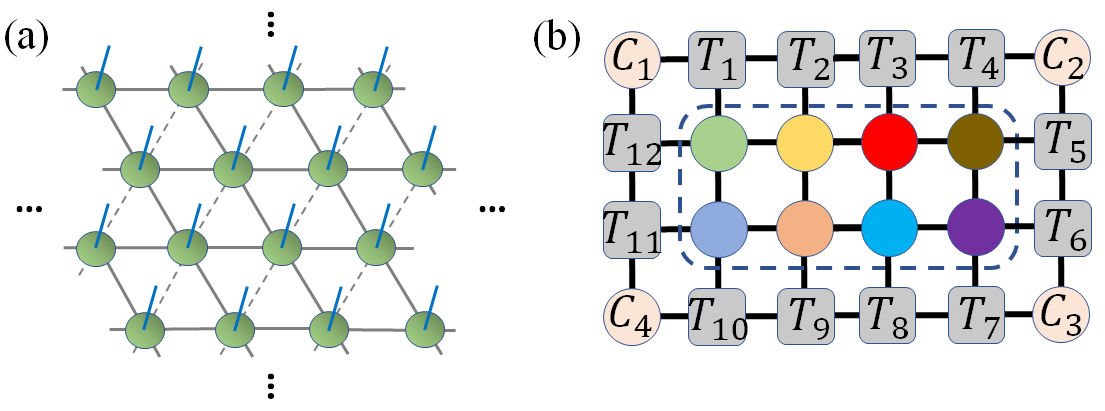}
\caption{(a) Composition of the rank-5 tensors into the formation of triangular lattice. Grey solid bonds represent the auxiliary bonds with dimension $D$, and blue open legs are the local indices with dimension $d=2$ for the $S=1/2$ system. Dashed bonds are not the real legs for the tensors but we act the $H_{ij}=\hat{S}^+_{i}\hat{S}^-_{j}+\hat{S}^-_i\hat{S}^+_j$ on these bonds too for simulating the triangular lattice. (b) The double layered tensors with bond dimension $D^2$ in the bulk, enclosed by the blue dashed box, and the environment tensors. The dimension of bonds interconnecting $C$ and $T$ tensors is labeled by $\chi$, with $\chi\ge D^2$ to guarantee the accuracy.}
\label{fig6}
\end{figure}

To study our effective model when the fluctuation is strong, we have no other option but to turn to the assistance of numerical tool. 
Due to the frustration that hinders the usage of quantum Monte Carlo, we adopt the infinite projected entangled-pair state~(iPEPS)~\cite{PhysRevLett.101.250602, NatRevPhys.1.538, RevModPhys.93.045003, MOTOYAMA2022108437} for our purpose.
There are two parts for iPEPS tensor network. One is the tensors within the repeating unit cell, where each composing unit is a rank-5 tensor with four auxiliary bonds~(dimension $D$) and one physical bond representing the size of local Hilbert space~(dimension $d=2$ for $S=1/2$).
Since originally iPEPS was designed for the square lattice, in the triangular lattice the coordinate number for each site is equal to six.
Instead of increasing the number of auxiliary bonds which would make the computation cumbersome, however, we act the local Hamiltonian also along the virtual bonds~(dashed bonds in Fig.~\ref{fig6}(a)) when calculating the energy.
As a result, the overall lattice structure is effectively equivalent to the triangular lattice.

Another important part for iPEPS lies on the environment tensors. Since the thermodynamics limit can be extrapolated through the corner transfer matrix renormalization group~(CTMRG) procedure~\cite{doi:10.1143/JPSJ.65.891, PhysRevB.80.094403, PhysRevLett.113.046402}, we need to first trace out the physical legs of each tensor with its adjoint one, forming the so-called double layered tensor where each bond is of dimension $D^2$.
We then adopt the CTMRG and after the fixed-point tensors have been found, they act as the effective ``bath" surrounding the bulk tensors, where we label the ones on the corners with $C$ and ones on the edges with $T$.
One example with the bulk size equal to $4\times 2$ is provided in Fig.~\ref{fig6}(b).

At last, we need to optimize the tensors so that the overall network can represent the ground state ansatz of the target Hamiltonian.
Here we adopt the variational optimization with the usage of automatic differentiation, first introduced by Liao \textsl{et al.}~\cite{PhysRevX.9.031041}.
We adopt the widely applicable package, peps-torch~\cite{Hasik}, for our calculation.
For readers aiming for some further information, we refer to our previous works in Refs.~\onlinecite{Commun.Phys.5.130, PhysRevB.107.224406}.

\section{\label{AppenC}Cluster mean-field theory}

In this Appendix we briefly introduce the CMFT method that we use for numerically solving Eq.~(\ref{Hamiltonian}). For adopting this method, we first divide our Hamiltonian into two parts: $H_C$ within the chosen cluster, and $H_{\partial C}$ that contains the terms connecting the bulk to the environment on the boundary of the cluster. $H_C$ possesses the exact form of Eq.~(\ref{Hamiltonian}), and the mean-field decoupling only takes place in $H_{\partial C}$, written as
\begin{equation}
\begin{aligned}
H_{\partial C}=&-t{\sum_{ij }}'(b^\dagger_{i}\langle b_{j}\rangle+H.C.)+{\sum_{ij}}'V_{ij} n_i \langle n_j\rangle,
\end{aligned}
\label{HamiltonianCMFT}
\end{equation}
where the prime symbol indicates that this summation is between site $i$ on the boundary of the cluster and site $j$ connected to $i$ outside the cluster.
Our effective Hamiltonian is then written as $H_{\text{eff}} = H_C + H_{\partial C}$. Although we write down the most general form in Eq.~(\ref{HamiltonianCMFT}), in this work we apply CMFT for the case with only the nearest-neighbor coupling.
We then exactly diagonalize the effective Hamiltonian and obtain the ground state to calculate the mean-field parameters, $\langle b_j \rangle$ and $\langle n_j \rangle$, for the following iteration.
After the mean-field parameters converge to consistent values, our calculation reaches its self-consistent solution. In CMFT, exact diagonalizations are performed within the chosen cluster. Therefore, it is not expected to include the long-range entanglement beyond the cluster size.
To study this finite size effect and infer the phases in the thermodynamic limit, a common practice is to compute the order parameters at varying cluster sizes and extrapolate it to the infinity.

CMFT is considered to be more accurate than the regular single-site mean-field theory, for that it can capture the correct correlation within the selected cluster. 
Thus it is suitable for states whose correlation length is small, meaning that deep inside the ordered phase~(or away from transition boundary) we can use CMFT and obtain quite accurate results. 
Because of the above-mentioned reason, for the boundary lines shown in Fig.~\ref{fig4}, we have extrapolated the energies and order parameters for both the first- and second-order transition boundaries to give a better outcome.
In addition, due to the randomness from the initialization process, CMFT could converge to distinct phases with the same set of parameters.
In this case, one should compare the energies among them to determine the correct ground state. Accordingly, it becomes a challenge when two stable phases appear with energy differences less than or comparable to the finite size effect.

In our CMFT calculation for Eq.~(\ref{Hamiltonian}), it provides definitive results on most of the phases shown in Fig.~\ref{fig4}. 
However, it is nontrivial to resolve the ground state phase between the stripe and kinked supersolid due to the fact that the energy difference between these two lies in the forth-order energy correction from the perturbation theory. 
For the $4\times 4$ cluster we employed, this difference is comparable to the finite size effect coming from the boundary, as can be estimated by investigating the changes in energy upon displacements of the solid patterns. 
However, by checking the order parameters we can still diagnose different phases.

\bibliography{draft}

\end{document}